\documentclass[12pt]{aipproc}
\layoutstyle{6x9}

\begin{document}

\title{Effective theories and constraints on new physics}


\author{R. Martinez, N. Poveda and J-Alexis Rodriguez}{
  address={Departamento de F\'{\i}sica, Universidad Nacional de Colombia\\
Bogota, Colombia}}

\copyrightyear  {2001}

\begin{abstract}

Anomalous  moments of the top quark arises from one loop
corrections to the vertices $\bar t t g$ and $\bar t t \gamma$. We study these anomalous couplings in  
different frameworks: effective theories, Standard Model and 2HDM. We use available experimental 
results in order to get bounds on these anomalous couplings.

\end{abstract}

\maketitle

\section{Introduction}

The top quark is the heaviest fermion in the standard model (SM) with a mass of
$174.3 \pm 5.1 $ GeV \cite{top}. In the framework of the SM, the couplings of the
top quark are fixed by the gauge symmetry. Anomalous couplings  between
the 
top quark and gauge bosons might affect the 
top quark production at high energies and also its decay rate. 
Precisely measured quantities with virtual top quark
contributions will yield further information 
regarding these couplings.

Since the top quark mass is so heavy, it is expected that its physics 
may be different from the lighter fermions and that the top quark
might couple quite strongly to the electroweak symmetry breaking sector. 
This suggests the SM is just an effective theory and that
the physics beyond the SM may be manifested through an effective Lagrangian 
involving the top quark.

The framework of effective theories, as a mean to parametrize physics beyond the
SM in a model independent way, has been used recently. Two cases have been 
consider in the literature, the decoupling case, which includes the Higgs boson, and
the non-decoupling case, where there is not any Higgs boson. We shall consider only
the first case, in which the SM is a low-energy limit of a renormalizable
theory.
In this approach, the effective theory parametrizes the effects at
low energy of the full renormalizable theory by means 
of high order dimensional non-renormalizable operators \cite{georgi}.

The effective Lagrangian approach is a convenient model independent
parametrization of the low-energy effects of the new physics
that may show up at high energies. Effective Lagrangians, employed to study
processes at a typical energy scale $E$ can be written as a power series in 
$1/\Lambda$, where the scale $\Lambda$ is associated with the heavy particles
masses of the underlying theory \cite{london}. The coefficients of the different
terms in the effective Lagrangian arise from integrating out the heavy
degrees of freedom. In order to define an effective Lagrangian it is necessary to specify the
symmetry and the particle content of the low-energy theory. In our case, we
require the effective Lagrangian to be $CP$-conserving, invariant under SM
symmetry $SU(2)_L\otimes U(1)_Y$, and to have as fundamental fields the
same ones appearing in the SM spectrum. Therefore we consider a Lagrangian
in the form
\begin{equation}
{\cal L}_{eff}={\cal L}_{SM}+\sum_{n}\alpha_n {\cal O}^n
\end{equation}
where the operators ${\cal O}^n$ are of dimension greater than
four.

\section{The anomalous magnetic moment}

The aim now is to extract indirect information on the
magnetic dipole moment of the top quark from LEP data, specifically we use
the ratios $R_{b}$ and $R_{l}$ defined by 
\begin{equation}
R_{b}=\frac{\Gamma (Z\rightarrow b\bar{b})}{\Gamma (Z\rightarrow hadron)}%
\;,\qquad R_{l}=\frac{\Gamma (Z\rightarrow hadron)}{\Gamma (Z\rightarrow l%
\bar{l})}, \; \Gamma_Z  \label{rbl}
\end{equation}
in the context of an effective Lagrangian approach. The
oblique and QCD corrections to the $b$ quark and hadronic $Z$ decay widths
cancel off in the ratio $R_{b}$. This property makes $R_{b}$ very sensitive
to direct corrections to the $Zb\overline{b}$ vertex, specially those
involving the heavy top quark , while $\Gamma _{Z}$ and $R_{l}$
are more sensitive to the oblique corrections.

In the present work, we consider the
following dimension six and $CP$-conserving operators, 
\begin{equation}
O_{uW}^{ab}=\bar{Q}_{L}^{a}\sigma ^{\mu \nu }W_{\mu \nu }^{i}\tau ^{i}\tilde{%
\phi}U_{R}^{b}\;,\qquad O_{uB}^{ab}=\bar{Q}_{L}^{a}\sigma ^{\mu \nu }YB_{\mu
\nu }\tilde{\phi}U_{R}^{b}\;,  \label{operador}
\end{equation}
where $Q_{L}^{a}$ is the quark isodoublet, $U_{R}^{b}$ is the up quark
isosinglet, $a$ , $b$ are the family indices, $B_{\mu \nu }$ and $W_{\mu \nu
}$ are the $U(1)_{Y}$ and $SU(2)_{L}$ field strengths, respectively, and $%
\tilde{\phi}=i\tau _{2}\phi ^{\ast }$. We use the notation introduced by
Buchm\"{u}ller and Wyler \cite{wyler}.
After spontaneous symmetry breaking, these fermionic operators generate also
effective vertices proportional to the anomalous magnetic moments of quarks.
The above operators for the third family give rise to the anomalous $t\bar{t}%
\gamma $ vertex and the unknown coefficients $\epsilon _{uB}^{ab}$ and $%
\epsilon _{uW}^{ab}$ are related with the anomalous magnetic
moment of the top quark through 
\begin{equation}
\delta \kappa _{t}=-\sqrt{2}{\frac{m_{t}}{m_{W}}}\frac{g}{eQ_{t}}%
(s_{W}\epsilon _{uW}^{33}+c_{W}\epsilon _{uB}^{33})\; ,
\end{equation}
where $s_{W}$ denotes the sine of the weak mixing angle.

The expression for $R_{b}$ is given by 
\begin{equation}
R_{b}=R_{b}^{SM}(1+(1-R_{b}^{SM})\delta _{b})\;,
\end{equation}
where $R_{b}^{SM}$ is the value predicted by the SM and $\delta _{b}$ is the
factor which contains the new physics contribution, and it is defined as
follows 
\begin{equation}
\delta _{b}={\frac{2\left( {g_{V}^{SM}g_{V}^{NP}+g_{A}^{SM}g_{A}^{NP}}%
\right) {+}\left( {g_{V}^{NP}}\right) ^{2}{+}\left( {g_{A}^{NP}}\right) ^{2}%
}{{(g_{V}^{SM})^{2}+(g_{A}^{SM})^{2}}}}
\end{equation}
and $g_{V}^{SM}$ and $g_{A}^{SM}$ are the vector and axial vector couplings
of the $Zb\bar{b}$ vertex. The contributions from new physics, eq. (\ref
{operador}), to $R_{l}$ and $\Gamma _{Z}$ are of two classes. One from
vertex correction to $Zb\bar{b}$ in the $\Gamma _{hadr}$ and the other from
the oblique correction through $\Delta \kappa $ in the $\sin ^{2}\theta _{W}$. 
These can be written as 
\begin{eqnarray}
R_{l} &=&R_{l}^{SM}(1-0.1851\;\Delta \kappa +0.2157\;\delta _{b})\;, 
\nonumber \\
R_{b} &=&R_{b}^{SM}(1-0.03\;\Delta \kappa +0.7843\;\delta _{b})\;,  \nonumber
\\
\Gamma _{Z} &=&\Gamma _{Z}^{SM}(1-0.2351\;\Delta \kappa +0.1506\;\delta _{b})
\end{eqnarray}
where $\Delta \rho $ is equal to zero for the operators that we are
considering.

We will consider the one loop contribution of the above effective operators to the $Zb\overline{b}$
vertex. After evaluating the Feynman diagrams, with
insertions of the effective operators $O_{uB}^{ab}$ and $O_{uW}^{ab}$ we
obtain 
\begin{eqnarray}
g_{V}^{NP} &=&4\sqrt{2}\epsilon _{uW}^{33}G_{F}m_{W}^{3}m_{t}\big \{3c_{W}(%
\tilde{C}_{12}-\tilde{C}_{11})-{\frac{m_{t}^{2}}{\sqrt{2}m_{W}^{2}}}%
(C_{12}-C_{11}+C_{0})  \nonumber \\
&&+{\frac{(1+a)}{8c_{W}}}(C_{11}+C_{12}+C_{0})+\frac{1}{\sqrt{2}}%
(C_{12}-C_{11}-C_{0})  \nonumber \\
&&-\frac{3a}{4c_{W}m_{Z}^{2}}(B_{1}-B_{0})\big \} , \nonumber \\
g_{A}^{NP} &=&4\sqrt{2}\epsilon _{uW}^{33}G_{f}m_{W}^{3}m_{t}\big \{-{\frac{a%
}{2c_{W}}}(C_{0}+C_{12}-C_{11})\nonumber \\
&-&\frac{1}{\sqrt{2}}(C_{12}-C_{0}-C_{11}) 
+\frac{m_{t}^{2}}{\sqrt{2}m_{W}^{2}}(C_{12}-C_{11}+C_{0})\nonumber \\
&&-{\frac{2m_{t}s_{W}^{2}}{m_{W}}}(\tilde{C}_{0}+\tilde{C}_{12}-\tilde{C}_{11}) 
-\frac{3}{4c_{W}m_{Z}2}(B_{1}-B_{0})\big \}\;
\end{eqnarray}
for the operator $O_{uW}$ and, 
\begin{eqnarray}
g_{V}^{NP} &=&g_{A}^{NP}=\frac{4\sqrt{2}}{3}\epsilon
_{uB}^{33}G_{F}m_{W}^{3}m_{t}\frac{s_{W}}{c_{W}}[\frac{m_{t}^{2}}{\sqrt{2}%
m_{W}^{2}}(C_{12}-C_{11}+C_{0})  \nonumber \\
&&-\frac{1}{\sqrt{2}}(-C_{11}+C_{12}-C_{0})]\;,
\end{eqnarray}
for the  $O_{uB}$. In the above equations $a=1-{\frac{8}{3}}%
s_{W}^{2} $ while $C_{ij}=C_{ij}(m_{W},m_{t},m_{t})$, $\tilde{C}_{ij}=\tilde{%
C}_{ij}(m_{t},m_{W},m_{W})$ and $B_{i}=B_{i}(0,m_{t},m_{W})$ are the
Passarino-Veltman scalar integral functions \cite{veltman}. The combination $%
B_{0}-B_{1}$ has a pole in $d=4$ dimensions that is identified with the
logarithmic dependence and can be replaced by $\ln \Lambda ^{2}/m_{Z}^{2}$.

Now we have various posibilities to explore the space of the parameters $%
\varepsilon _{uW}^{33}$, $\varepsilon _{uB}^{33}$ and $\delta \kappa _{t}$. Using experimental values
for $\Gamma_Z$, $R_b$ and $R_l$ we find 
the allowed region in the plane $\varepsilon _{uW}^{33}-$ $\varepsilon
_{uB}^{33}.$ If we do not neglect the term of the order $\left(
g^{NP}\right) ^{2}$, we get the following expressions 

\begin{eqnarray}
-0.057\varepsilon _{B}+0.058\varepsilon _{B}^{2}+0.053\varepsilon
_{W}+0.0015\varepsilon _{W}^{2}+0.01\varepsilon _{B}\varepsilon _{W}
&=&1-(\Gamma _{Z}^{\exp }/\Gamma _{Z}^{SM}),  \nonumber \\
-0.023\varepsilon _{B}+0.026\varepsilon _{B}^{2}+0.016\varepsilon
_{W}+0.0007\varepsilon _{W}^{2}+0.0048\varepsilon _{B}\varepsilon _{W}
&=&1-(R_{b}^{\exp }/R_{b}^{SM}),  \nonumber \\
-0.656\varepsilon _{B}+0.686\varepsilon _{B}^{2}+0.536\varepsilon
_{W}+0.018\varepsilon _{W}^{2}+0.1264\varepsilon _{B}\varepsilon _{W} 
&=&1-(R_{l}^{\exp }/R_{l}^{SM}), \nonumber
\end{eqnarray}
where we have omitted the superindex $33$. 

The SM
values for the parameters that we have used are 
\begin{eqnarray}
\Gamma _{Z}&=&2.4963 GeV,\;\; R_{l}=20.743, \nonumber \\
R_{b}&=&0.21572, \Gamma _{hadr}=17427 MeV, \;\; \Gamma_{l}=84.018 MeV,\nonumber  
\end{eqnarray}
with the input parameters: 
\begin{eqnarray}
m_{t}&=&174.3 GeV, \;\; \alpha
_{s}(m_{Z})=0.118, \nonumber \\
m_{Z}&=&91.1861 GeV, \;\; m_{H}=100 GeV,\;\; \Lambda =1
TeV\nonumber . 
\end{eqnarray}
And the experimental values are 
\begin{eqnarray}
\Gamma _{Z}&=&2.4952\pm 0.0023 GeV, \;\;
R_{l}=20.804\pm 0.050, \nonumber \\
R_{b}&=&0.21653\pm 0.00069 \nonumber.
\end{eqnarray}

After doing a $\chi^2$ analysis at $95\%$ C.L. we find the allow region for $\varepsilon _{uW}^{33}-$ $\varepsilon
_{uB}^{33}$ parameters. In this kind of scenarios new
physics is explored assuming that its effects are smaller than the SM
effects, consequently one expect that $\left|
g_{V,A}^{NP}/g_{V,A}^{SM}\right| \ll 1$ and, then we get  
the inequalities $\left| \varepsilon
_{uW}^{33}\right| \leq 0.11,$ $\left| \varepsilon _{uB}^{33}\right| \leq
0.48$. By using the eq. $(4)$ and the bounds got
in the numerical analysis, we obtain for $\delta \kappa _{t}$ the following values 
\begin{eqnarray}
-2.94\leq &\delta\kappa _{t}&\leq 1.3, \; -0.76\leq \delta \kappa _{t}\leq 1.9 ,\nonumber \\
 -1.3\leq &\delta\kappa _{t}&\leq 1.7
\end{eqnarray}
which correspond to $\Gamma
_{Z},\ $\ $R_{b}$ and $R_{l}$, respectively. Therefore for these observables 
$\Gamma _{Z},\ $\ $R_{b}$ and $R_{l}$, we get the allowed region 
\begin{equation}
-2.94\leq\delta \kappa_t \leq 1.9.
\end{equation}

\section{Anomalous chromomagnetic dipole moment I}

We are interested in studying possible deviations
from the SM on the decay $b\to s\gamma$ within the context of the effective
Lagrangian approach. Several authors have been used the CLEO 
results on radiative 
$B$ decays to set bounds on the anomalous coupling of the
t-quark.
We will use dimension-six operator
which are full strong and electroweak gauge invariant and contribute to 
$b\to s\gamma$ in order to bound the chromomagnetic dipole moment of
the top quark.

Since the anomalous chromomagnetic dipole moment of the top quark appears in
the top quark cross section, it is possible, due to uncertainties,  
to estimate the constraints that it would impose on the $\Delta \kappa_g^t$.
For the LHC, the
anomalous coupling is constrained to lie in the range $-0.09 \leq \Delta
\kappa_g^t \leq 0.1$ \cite{atwood}. Similar range is obtained for the future NLC.  The influence of an anomalous $\Delta \kappa$ on the cross section
and associated gluon jet energy for $t \bar t g$ has 
been also analyzed. Events produced at 
$500$ GeV in $e^+ e^-$ colliders, with a cut on the gluon energy
of $500$ GeV and integrated luminosity of $30 fb^{-1}$,  lead to a
bound of $-2.1 \leq \Delta \kappa_g^t \leq 0.6$ \cite{atwood}.

>From the experimental information it is possible to get a
limit on the $\Delta \kappa_g^t$ from Tevatron. 
Following the reference by F. del Aguila \cite{aguila} and  assuming that the only non-zero coupling is precisely the chromomagnetic dipole moment of 
the top quark, we find from the collected data that the allowed region is $\vert  \Delta \kappa_g^t \vert \leq 0.45$.

We consider the following dimension-six, CP-conserving
operators, which are $SU(3)_C\otimes S(2)_L\otimes U(1)_Y$ gauge invariant
\begin{equation}
{\cal O}^{ab}_{uG}=\bar{Q}_L^a \sigma_{\mu\nu} G^{\mu\nu\; i}
\frac{\lambda^i}{2}\tilde{\phi} u_R^b \; ,
\label{six}
\end{equation}
where $G^{\mu\nu\; i}$ is the gluon field strength tensor and $a, b$ are
the
family indices. The above operator gives rise to the anomalous $t\bar{t}g$ 
vertex and its respective 
unknown coefficient $\epsilon_{ab}^{uG}$ is related with the anomalous
chromomagnetic 
moment of the top quark by
\begin{equation}
\delta\kappa^t_g =\sqrt{2}\frac{g}{g_s}\frac{m_t}{M_W}\epsilon_{uG}^{33} \; .
\end{equation}
where by means of the dimension 5 coupling to 
an on-shell gluon, the anomalous chromomagnetic dipole moment of the top quark is defined as
\begin{equation}
L_5=i (\frac {\Delta \kappa_g^t}{2}) \frac{g_s}{2 m_t} \bar{u}(t) 
\sigma_{\mu \nu} q^\nu T^a u(t) G^{\mu ,a}
\end{equation}
with $g_s$ and $T^a$ are the $SU(3)_c$ coupling and generators,
respectively.

The effective Hamiltonian used to compute the $b\to s$ transition is given
by \cite{buras}
\begin{equation}
H_{eff}=-\frac{4 G_F}{\sqrt{2}} V_{ts}^{*} V_{tb} \sum_{i=1}^{8} c_i(\mu)
{\cal O}_i(\mu) \; ,
\end{equation}
where $\mu$ is the energy scale at which $H_{eff}$ is applied. For
$i=1 - 6$, ${\cal O}_i(\mu)$ correspond to four-quark operators, ${\cal
O}_7(\mu)$
is the electromagnetic dipole moment and ${\cal O}_8(\mu)$ is the chromomagnetic
dipole operator. At low energy, $\mu\approx m_b$, the only operator that
contributes
to $b\to s$ transition is ${\cal O}_7(\mu)$ which results from a mixing
among the ${\cal O}_2(M_W)$,
${\cal O}_7(M_W)$ and ${\cal O}_8(M_W)$ operators.

The total contribution of the effective operator (\ref{six}) to
the ${\cal O}_8(M_W)$ operator can be written as:
\begin{equation}
c_8(M_W)=c_8(M_W)^{SM}+\delta\kappa_g^t \Delta c_8(M_W) \; ,
\end{equation}
where
\begin{eqnarray}
\Delta c_8(M_W)&=& \frac{1}{4 V_{ts}}
\ln\left(\frac{\Lambda^2}{M_W^2}\right)
+\frac{1}{V_{ts}}\frac{x-x^2+x (2-x)\ln(x)}{8(1-x)^2} \nonumber \\
&+&\frac{3 x - 4 x^2 + x^3 + 2 x\ln(x)}{8 (1-x)^3}  
\end{eqnarray}
and $x=m_t^2/M_W^2$.

Using the recent data from CLEO collaboration for the
branching fraction of the process $B(b \to s
\gamma)=(3.21 \pm 0.43 \pm 0.27) \times 10^{-4}$ \cite{cleo}, we  get an allowed region for the
anomalous chromomagnetic dipole moment of the top quark to be \cite{nos} 
\begin{equation}
-0.03 \leq \Delta\kappa_g^t \leq 0.01
\end{equation}

\section{ Anomalous chromomagnetic dipole moment II}

Our next objective  is to evaluate the contribution 
at the one loop-level to the anomalous
chromomagnetic dipole moment of the top quark in different scenarios with the
gluon boson on-shell. Beginning with the SM, the typical QCD correction through gluon exchange
implies two different Feynman diagrams. After the explicit calculation of the 
loops, we find that 
\begin{equation}
\Delta \kappa_g^t=-\frac 16 \frac{\alpha_s(m_t)}{\pi} .
\end{equation} 
We note that its
natural size is of the order of $\alpha_s/ \pi$ similar to the QED anomalous
coupling, but now 
in combination with a factor $-1/6$ coming from the color structure in the
diagram i.e.  $T^a T^b T^a=-T^b/6$ with $T^a$ 
being the generators of $SU(3)_C$.

The other possible contribution in the framework of the SM comes
from electroweak interactions. The relevant contributions occur when neutral
Higgs boson and the would-be Goldstone boson of $Z$ are involved in the loop.
This contribution reads 
\begin{equation}
  \Delta \kappa_g^t= - \frac{\sqrt{2} G_F m_t^2}{8
\pi^2} [H_1(m_h)+H_2(m_Z)] ,  
\end{equation}
where
\begin{eqnarray}
H_1(m)&=& \int_0^1 dx \frac{x-x^3}{x^2-(2-m^2/m_t^2)x+1}  ,\nonumber \\
H_2(m)&=& \int_0^1 dx \frac{-x+2x^2-x^3}{x^2-(2-m^2/m_t^2)x+1}. \nonumber
\end{eqnarray}
The SM contribution is showed in figure 1 where we have added the QCD
contribution. 
It is worth noting that the behaviour of the curve for a large Higgs
boson mass indicates decoupling and 
that the values of $\Delta \kappa_g^t$  lie within 
the allowed region for $\Delta \kappa_g^t$ coming from $b \to s \gamma$.

The contributions within a general 2HDM will be different
from the SM contributions because of the presence of the virtual five
physical Higgs bosons which appear in any two Higgs doublet model after
spontaneous symmetry breaking: $H^0$, $A^0$, $h^0$, $H^\pm$ \cite{doshiggs}.
Therefore, 2HDM predictions depend on their massses and
on the two mixing angles $\alpha$ and $\beta$. For small $\beta$, the charged
Higgs boson contribution is suppressed due to 
its large mass  and the small bottom quark mass.

The expression  for the contribution of the neutral Higgs bosons is given by 
\begin{eqnarray} 
\Delta \kappa_g^t&=&\frac{\sqrt{2} G_F}{8 \pi^2} 
[\lambda_{H^0 tt}^2 H_1(M_H^2)+\lambda_{h^0 tt}^2 H_1(M_h^2)  \nonumber \\
&+&\lambda_{A^0tt}^2 H_2(M_A^2)+\lambda_{G^0 tt}^2 H_2(M_Z^2)]
\end{eqnarray}
where $\lambda_{itt}$ are the Yukawa couplings in the so-called models 
of type I,
II and III . Table I shows the couplings in the usual convention.
\begin{table}[htbp]
\begin{tabular}{l c c c }\\  
\hline
$\lambda_{itt}$ & model type I & model type II & model type III \\ \hline
$\lambda_{H^0 tt}$ & $\frac{m_t \sin \alpha}{\sin \beta}$ & $\frac{m_t \cos
\alpha}{\cos \beta}$ & $(1+\frac{\eta^U_{tt}}{\sqrt{2}})m_t \sin \alpha$\\ 
$\lambda_{h^0 tt}$ & $\frac{m_t \cos \alpha}{\sin \beta}$ & $\frac{m_t \sin
\alpha}{\cos \beta}$ & $(1+\frac{\eta^U_{tt}}{\sqrt{2}})m_t \cos \alpha$\\ 
$\lambda_{A^0tt}$ & $\cot \beta m_t$ & $\tan \beta m_t$ & $\frac{\eta^U_{tt}
m_t}{\sqrt{2}}$ \\ 
$\lambda_{G^0tt}$ & $m_t$ & $m_t$ & $m_t$ \\
\hline\\
\end{tabular}
\caption{Couplings of the Higgs eigenstates with the top quark in 2HDM. We
omit the factor $g/2 m_W$ and in the model type III we use the Sher-Cheng
approach for the flavour changing couplings.}
\end{table}

The Yukawa couplings of a given fermion to the Higgs scalars are proportional
to the mass of the fermion and they are therefore naturally enhanced in this
case. In the model type III appears flavour changing neutral couplings at tree level which can be parametrized in the Sher-Cheng approach where a natural value for 
the flavour changing couplings from different families should be of the order of the 
geometric average of their Yukawa couplings, $h_{ij}=g \eta_{ij}\sqrt{m_i m_j}/(2 m_W)$ with $\eta_{ij}$ of the order of one.

\begin{figure}
  \includegraphics[angle=0,width=6cm]{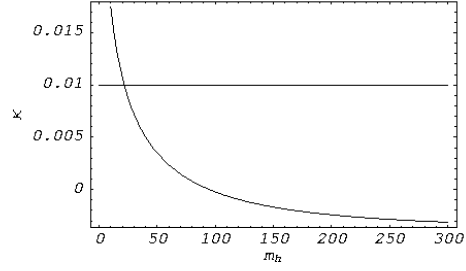}
  \caption{Standar Model contribution to the anomalous cromomagnetic dipole moment
of the top quark vs the higgs boson mass}
\end{figure}

In order to show the behaviour of the  contribution of the 2HDM to the anomalous chromomagnetic dipole moment of the top quark, we 
evaluate explicitly the contribution for couplings type II. We show in figure 2 the allowed region (above the curve)  for
the plane $\tan \beta$ vs $m_H$ using equation (21) and assuming that  $-0.03 \leq \Delta \kappa_g^t \leq  0.01$  from $b \to s \gamma$.  We fix the following parameters: $m_H=m_h$,  $m_A=200(400)$ GeV solid line (dashed line). The solid line for the scalar Higgs mass smaller (bigger) than $240$ GeV corresponds to the cut between equation (21) and the upper(lower) limit from $b \to s \gamma$. 

\begin{figure}
  \includegraphics[angle=0,width=6cm]{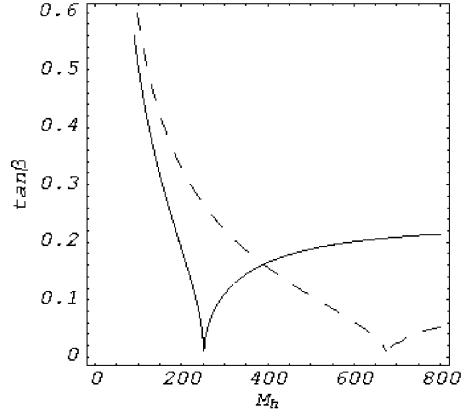}
  \caption{Contour plot for the contribution of 2HDM to the anomalous cromomagnetic dipole moment 
of the top quark in the plane $\tan\beta - m_H$ for $M_A=200$ GeV(solid line) and $M_A=400$ GeV (dashed line)
using the bound from the $b\to s\gamma$ process. The allowed region is above the curve}
\end{figure}

\begin{theacknowledgments}
We thank COLCIENCIAS for finantial support
\end{theacknowledgments}


          {\ifthenelse{\equal{\AIPcitestyleselect}{num}}
             {\bibliographystyle{arlonum}}
             {\bibliographystyle{arlobib}}

\begin{thebibliography}{99}
\bibitem{top}K. Hagiwara {\em et.al.\ }(Particle Data Group), \ Phys. Rev. D 66, 1 (2002).
\bibitem{georgi}H. Georgi, Nucl. Phys. B 361, 339 (1991); A. De Rujula {\em et.al.\ }, Nucl. Phys. B 369, 3 (1992); 
J. Wudka, Int. J. Mod. Phys. A 9, 2301 (1994).
\bibitem{london}C. P. Burgess and D. London, Phys. Rev. D 48, 4337 (1993); Phys. Rev. Lett. 69, 3428 (1993).
\bibitem{wyler}W. Buchmuller and D. Wyler, Nucl. Phys. B 268, 621 (1986); Phys. Lett. B 197, 379 (1987).
\bibitem{veltman}G. Passarino and M. Veltman, Nucl. Phys. B 160, 151 (1979)
\bibitem{atwood}D. Atwood, A. kagan and T. Rizzo, Phys. Rev. D 52, 6264 (1995); T. G. Rizzo, hep-ph/9902273; T. Rizzo, Phys. Rev. D 50, 4478 (1994).
\bibitem{aguila}F. de Aguila, hep-ph/9911399.
\bibitem{buras}A. J. Buras {\em et.al.\ }, Nucl. Phys. B 424, 379 (1994); M. Misiak, Phys. Lett. B 269, 161 (1991).
\bibitem{cleo}CLEO collaboration, hep-ex/0108032.
\bibitem{nos}R. martinez and J-Alexis Rodriguez, Phys. Rev. D 55,3212 (1997).
\bibitem{doshiggs}J. Gunion, H. Haber, G. Kane and S. Dawson, The Higgs Hunter's Guide, Frontiers in Physics. Edt. D. Pine (1990). 
\end{thebibliography}
          }

\end{document}